\documentstyle[aps,prl,floats,epsf,twocolumn]{revtex}

\newcommand{\xdir}{$[ 1\overline{1} 0 ]$}
\newcommand{\ydir}{$[ 110 ]$}
\newcommand{\zdir}{$[ 001 ]$}

\newcommand{\Bpar}{$B_{\parallel}$}

\newcommand{\Rxx}{$R_{xx}$}
\newcommand{\Ryy}{$R_{yy}$}

\newcommand{\Rhard}{$R_{hard}$}
\newcommand{\Reasy}{$R_{easy}$}
\newcommand{\rhohard}{$\rho_{hard}$}
\newcommand{\rhoeasy}{$\rho_{easy}$}

\begin{document}

\draft
\wideabs{
\title{The Onset of Anisotropic Transport of Two-Dimensional Electrons in High 
Landau Levels: An Isotropic-to-Nematic Liquid Crystal Phase Transition?}

\author{K.~B. Cooper$^1$, M.~P. Lilly$^{1}$\cite{lillyaddress}, J.~P. Eisenstein$^1$, L.~N. 
Pfeiffer$^2$, and K. W. West$^2$}
\address{$^1$California Institute of Technology, Pasadena CA 91125 \\
         $^2$Bell Laboratories, Lucent Technologies, Murray Hill, NJ 07974}

\maketitle

\begin{abstract}
The recently discovered anisotropy of the longitudinal resistance of two-dimensional 
electrons near half filling of high Landau levels is found to 
persist to much higher temperatures $T$ when a large in-plane magnetic field 
\Bpar\ is applied.  Under these conditions we find that the longitudinal 
resistivity scales quasi-linearly with \Bpar/$T$. These observations 
support the notion that the onset of anisotropy at \Bpar$=0$ does not reflect 
the spontaneous development of charge density modulations but may instead signal 
an isotropic-to-nematic liquid crystal phase transition. 
\end{abstract}

}

Transport experiments\cite{lilly1,du} on clean
two-dimensional electron systems (2DES) at high perpendicular magnetic
field have recently revealed a new class of collective
phases at low temperatures. These new phases, which have been observed
only when electrons occupy the excited Landau levels, are quite
different from the fractional quantized Hall liquids found
almost exclusively in the lowest ($N=0$) Landau
level\cite{perspectives}. The new phases exhibit strong electrical
transport anisotropies near half filling and curious re-entrant
insulating behavior in the flanks of the third and higher ($N \ge 2$)
Landau level (LL). At the phenomenological level, these properties are
remarkably consistent with those expected from the ``stripe" and
``bubble" charge density waves (CDW) predicted on the basis of
Hartree-Fock (HF) theory\cite{KFS,MC}. These inhomogeneous states
are stabilized by an exchange energy-driven tendency to phase separate
which is particularly effective in the excited LLs. Near half filling,
the ground state is approximately a unidirectional CDW, i.e. a stripe phase.
The anisotropy of the longitudinal resistance observed at 
this filling is presumed to result from very different electrical conductivities
parallel and perpendicular to stripes whose orientation relative to the crystal
axes of the host semiconductor is determined by some, as yet unknown, symmetry-breaking
field.

With the filling factor $\nu$ defined as the ratio of the 2D electron
density $n_s$ to the degeneracy $eB/h$ of a single spin resolved LL, the
simplest HF stripe state consists of parallel strips in which
the filling factor $\nu$ alternates between adjacent integers. For
example, at {\it average} filling $\nu=9/2$, corresponding to half
filling of the lower spin branch of the N=2 LL, the local filling factor
alternates between $\nu=4$ and $\nu=5$. The period of this alternation
is set by the competition between the direct and exchange Coulomb
interactions and is estimated to be on the order of 100 nm in typical
samples. These same interactions also determine the temperature at which
the stripes form and HF estimates are in the few Kelvin
range\cite{KFS}. This contrasts with the experimental observation that
resistive anisotropy only sets in below about 100 mK. Although disorder
in the 2D system may account for some of this discrepancy, a more
interesting idea is that the onset of anisotropy does not in fact
signal the initial development of charge density modulation, but instead
reflects the orientational ordering of local regions having pre-existing
stripe order. This idea, first suggested by Fradkin and
Kivelson\cite{fradkin}, emerges naturally from the view that quantum and
thermal fluctuations destroy the long range translational order of the
stripe phases and render them analogous to nematic liquid crystals. The
onset of anisotropy is therefore viewed as an isotropic-to-nematic phase
transition. Wexler and Dorsey\cite{wexler}
have estimated the transition temperature to be on the order of 200 mK,
not far from the experimental result. In this paper, we report
temperature-dependent magneto-transport studies which address this
issue. Our results reveal that the sharp thermal onset of resistive
anisotropy is replaced by a heavily smeared transition when a strong
in-plane magnetic field component \Bpar\ is added to the existing
perpendicular field. This observation, which is reminiscent of
the behavior of a ferromagnet in a strong symmetry-breaking field,
supports the isotropic-to-nematic liquid crystal transition idea. In
addition, we report an intriguing scaling of the resistive anisotropy
with $T$/\Bpar\ which may prove important in developing a quantitative
understanding of the putative nematic phase.

The sample used for the present experiments is a conventional
single-interface GaAs/AlGaAs heterojunction grown by molecular beam
epitaxy (MBE) on an \zdir-oriented GaAs substrate. A thin sheet of Si
impurities is positioned in the ${\rm Al_{0.24}Ga_{0.76}As}$ 80 nm from
the interface with the GaAs. Charge transfer from these donors results
in a 2DES whose density and low temperature mobility are
$n_s=1.48\times10^{11}$ cm$^{-2}$ and $\mu=1.1\times10^7$ cm$^2/$Vs,
respectively. These parameters are obtained after brief low temperature
illumination of the sample by a red light emitting diode. The sample
itself is a 4x4 mm square cleaved from the parent wafer. The edges of
this square are parallel to the \ydir\ and \xdir\ crystallographic
directions. Eight diffused In ohmic contacts are positioned at the
corners and midpoints of the sides of the sample. Longitudinal
resistance measurements are performed by driving a 20 nA, 13 Hz
a.c.\ current between opposing midpoint contacts and detecting the
resulting voltage between corner contacts. We denote by \Rxx\ and
\Ryy\ the resistances obtained for average current flow parallel to
the \xdir\ and \ydir\ directions, respectively. Adjustable in-plane
and perpendicular magnetic fields are obtained by
$in~situ$ tilting of the sample relative to the field $B$ of a single
superconducting solenoid\cite{bpar}.

\begin{figure} 
\begin{center}
\epsfxsize=3.3in
\epsfclipon
\epsffile[16 4 353 257]{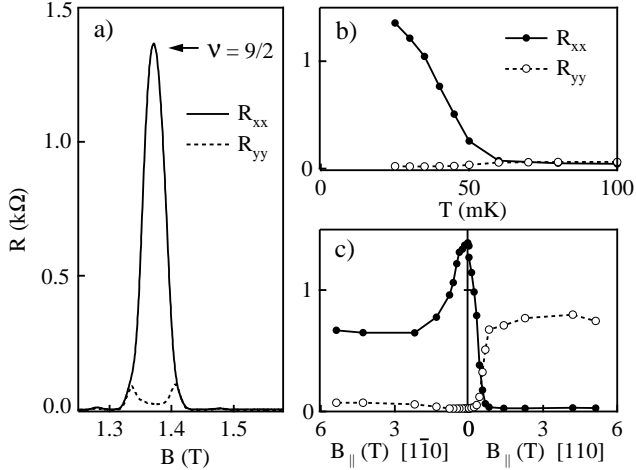}
\end{center}
\caption[figure 1]{a) Longitudinal resistance anisotropy
around $\nu=9/2$ at $T=25$ mK. Solid trace: \Rxx; average
current flow along \xdir\ crystallographic direction. Dashed trace:
\Ryy; average current flow along \ydir. b) Temperature
dependence of resistances at $\nu=9/2$. c) \Rxx\ and
\Ryy\ at $\nu=9/2$ at $T=25$ mK \emph{vs.} in-plane magnetic
field along \ydir\ and \xdir. 
}
\end{figure}

Figure 1a summarizes the resistance anisotropy effect observed
around filling factor $\nu=9/2$. The solid and
dashed curves represent \Rxx\ and \Ryy, respectively, at
$T=25$ mK in a purely perpendicular magnetic field (i.e.\ \Bpar$=0$). 
The anisotropy is largest at half-filling, where
\Rxx/\Ryy\ reaches a value of approximately 60 in this sample.
As Fig.~1b shows,
raising the temperature causes \Rxx\
to fall rapidly and \Ryy\ to rise somewhat. Above about 60 mK both
resistances are saturated at values less than 60 $\Omega$. This transport
anisotropy is also present at filling factors $\nu=11/2$, 13/2, 15/2,
etc., albeit with decreasing strength. As reported
previously\cite{lilly1,du}, no anisotropy is observed in the $N=1$ first
excited Landau level\cite{N1LL} nor in the $N=0$ lowest LL. 

\begin{figure} 
\begin{center}
\epsfxsize=3.3in
\epsfclipon
\epsffile[18 104 345 383]{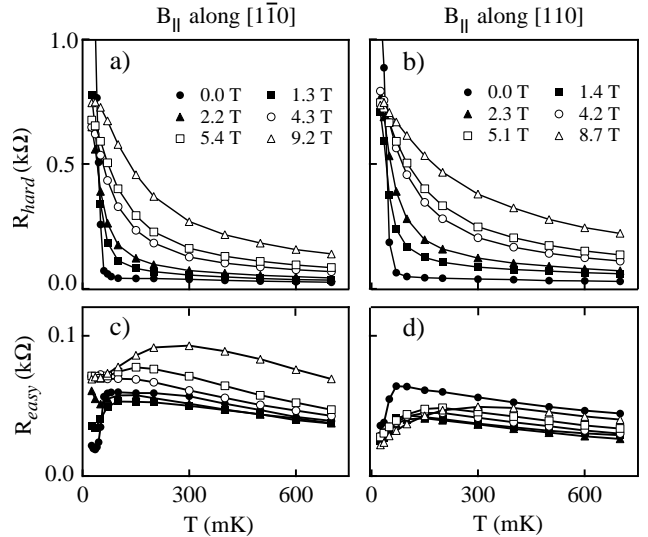}
\end{center}
\caption[figure 2]{Temperature dependencies of longitudinal resistance
in hard and easy directions at $\nu=9/2$.  a) and b): \Rhard\ \emph{vs.}
$T$ for various in-plane magnetic fields \Bpar\ directed along \xdir\ and \ydir,
respectively.  c) and d): Analogous results for \Reasy.  Note
magnified vertical scale.
}
\end{figure}

An in-plane magnetic field can alter the
orientation of the resistive anisotropy in high Landau
levels\cite{pan,lilly2}. Figure 1c summarizes this effect by displaying
the dependences of $R_{xx}$ and $R_{yy}$ at $\nu=9/2$ and $T=25$ mK
on \Bpar. In the right half of the panel \Bpar\ is
directed along the \ydir\ direction; in the left half it is along \xdir.
For \Bpar\ along \ydir, \Rxx\ quickly falls to a small value just
as \Ryy\ quickly rises to a large value: The in-plane field
interchanges the ``hard" and ``easy" transport directions.
On the other hand, with \Bpar\ directed along \xdir, no such
interchange occurs. These data demonstrate that large in-plane fields
can overwhelm the native symmetry-breaking potential in the sample and
force the hard resistance direction to be parallel to the in-plane
field. Jungwirth {\it et
al.}\cite{jungwirth} have argued that this effect arises from the finite
thickness of the 2DES. For samples such as the present one, their HF theory
reveals an energetic advantage for the stripes to lie perpendicular to
the in-plane magnetic field. By comparing the in-plane field required to
interchange the anisotropy axes with this theory, an estimate of the
native anisotropy energy may be obtained: Typical values are around 1 mK
per electron\cite{cooper}. 

Figures 2a and 2b display the temperature dependencies of the longitudinal
resistance in the ``hard'' and ``easy'' directions, \Rhard\ and \Reasy,
at $\nu=9/2$.  The data in Fig.~2a were
taken with the in-plane magnetic field directed along \xdir.  
In this case, the
in-plane field does not interchange the principal axes of the resistance
anisotropy and thus $R_{hard}=R_{xx}$. For Fig.~2b, however, the
in-plane field is along \ydir\ and therefore {\it does} interchange the
hard and easy directions, provided \Bpar\ $>0.5$ T. With the
exception of the \Bpar\ $=0$ data, the traces in Fig.~2b correspond to
\Ryy, which is \Rhard\ at the in-plane fields chosen. Figures 2c
and 2d show the corresponding results for \Reasy, the longitudinal
resistance in the easy direction, on a magnified scale. Note that \Reasy\
is small compared to \Rhard\ for all \Bpar\ at low temperatures.

It is clear from Figs.~2a and 2b that a large \Bpar\
systematically broadens out the variation of \Rhard\ at low
temperatures. At \Bpar\ $=0$ the bulk of the change in \Rhard\ occurs
in a relatively narrow window below $T\approx 60$ mK. In contrast,
at \Bpar\ $=9$ T \Rhard\ diminishes smoothly up to temperatures above
600 mK. This qualitative broadening effect due to the
in-plane magnetic field occurs for both orientations of the field
relative to the crystal directions. Interestingly, the effect is not
obviously present in \Reasy, although close inspection does reveal
some softening at low temperatures. 

The thermal broadening of the resistive anisotropy in high Landau levels
induced by strong in-plane magnetic fields is reminiscent of the
behavior of a mean-field ferromagnet in the presence of an external
magnetic field\cite{pathria}. Very weak external fields merely break
rotational symmetry and orient the magnetization of the system which
otherwise develops spontaneously below the Curie temperature $T_c$.
In the presence of a strong external field, however, the
ferromagnetic transition is broadened out in temperature and substantial
magnetization is present at temperatures $T>T_c$. For $T>>T_c$ the
magnetization is that of the free spins in the system. By analogy, our
observation that the high Landau level resistive anisotropy persists to
high temperatures in the presence of large \Bpar\ suggests that local
stripe ``moments'' exist in the 2DES at such temperatures. Lacking a
strong symmetry-breaking in-plane field, these moments are disordered at
high temperature and transport is isotropic. This scenario is consistent with
the nematic-to-isotropic liquid crystal transition proposed by Fradkin and
Kivelson\cite{fradkin}. 

To pursue this idea further we have extracted, in an
approximate manner, the microscopic {\it resistivities} of the 2DES from
the measured macroscopic {\it resistances}. As emphasized by
Simon\cite{simon}, the resistances \Rhard\ and \Reasy\ in square
samples exaggerate the actual anisotropy of the underlying resistivities
\rhohard\ and \rhoeasy\cite{hallbars}. This stems from the fact
that the current distribution in an anisotropic sample is quite
different from that in an isotropic one. In particular, currents flowing
in the easy direction are channeled in toward the axis connecting the
source and drain contacts. As a result, the voltages present at the
remote contacts used to determine \Reasy\ can be extremely small and
therefore particularly sensitive to local inhomogeneities and other
defects in the sample. Conversely, current flow in the hard direction is
spread out more uniformly across the sample and the voltages used to
determine \Rhard\ are robust.  As a consequence, we will focus our attention
on \rhohard.

\begin{figure} [t]
\begin{center}
\epsfxsize=3.3in
\epsfclipon
\epsffile[14 29 386 193]{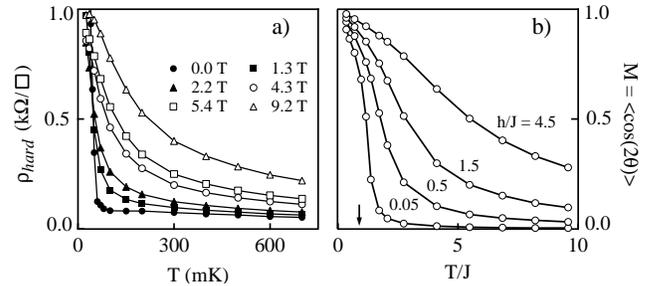}
\end{center}
\caption[figure 3]{a) Temperature dependence of \rhohard\ at $\nu=9/2$
for various \Bpar\ along \xdir. b) Temperature dependence of order parameter
of the 2D XY model for various symmetry-breaking fields $h/J$.  Arrow denotes
Kosterlitz-Thouless transition temperature for $h=0$.
}
\end{figure}

Figure 3a displays the temperature dependence of
\rhohard\ at $\nu=9/2$ for several \Bpar\ along \xdir. These
resistivities were computed from the measured resistances \Rhard\ and
\Reasy\ under the assumption of Simon's\cite{simon} classical model
of the current distribution in the sample. While this model has
shortcomings, its success in reconciling the differences between the
resistive anisotropy in square and Hall bar samples suggests that it
provides reasonable estimates of the
resistivities\cite{hallbars}. In any case, Fig. 3a reveals that
\rhohard\ exhibits the same broadening of its temperature
dependence as does \Rhard\ when a large in-plane magnetic field is
applied. 

Fradkin, {\it et al.}\cite{fradkin,fradkin2} and Wexler and
Dorsey\cite{wexler} have argued that the stripe system undergoes a
finite temperature Kosterlitz-Thouless (KT) transition from an isotropic
state possessing only local stripe order to a nematic phase whose
director field possesses algebraically decaying correlations. Below the
KT transition temperature, the system is singularly susceptible to
external symmetry breaking fields and is thus readily oriented by some
(as yet unexplained\cite{cooper}) weak symmetry breaking field in
the GaAs/AlGaAs sample. Fradkin, {\it et al.} have
modeled the system as a classical 2D XY magnet, modified to accomodate
a director field. The order parameter for the system is taken to be
$M=<$cos$(2\theta)>$, where $\theta$ is the angle between the director and
an arbitrarily weak symmetry-breaking field $h$. Using Monte Carlo
methods, they have been able to achieve good fits to the observed
temperature dependence of the resistive anisotropy at $\nu=9/2$ (in a
different sample) using a sensible value of the exchange energy, $J$,
and a weak symmetry breaking field, $h \sim 0.05J$. We have
extended these calculations to larger $h$ to simulate
the effect of the large in-plane magnetic field used in the present
experiments\cite{cooper2}. As Fig. 3b shows, the temperature dependence of $M$ is
systematically broadened out as $h$ increases and bears a strong
qualitative resemblance to the behavior of \rhohard\ (and
\Rhard) at large \Bpar. 

\begin{figure} [t]
\begin{center}
\epsfxsize=3.3in
\epsfclipon
\epsffile[14 18 346 385]{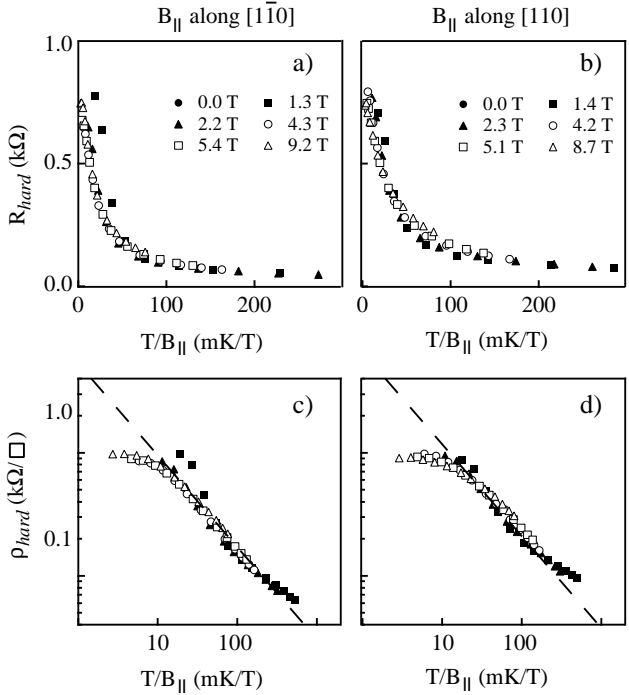}
\end{center}
\caption[figure 4]{
a) and b): \Rhard\ $vs.$ the scaled temperature $T/$\Bpar\ for
\Bpar\ directed along \xdir\ and \ydir, respectively.
c) and d): Log-log plot of the hard axis resistivity, \rhohard,
calculated from the measured resistances, {\it vs.}
$T/$\Bpar. Dashed lines are power laws: 
\rhohard$\propto (T/$\Bpar$)^{-\alpha}$ with $\alpha=0.75$.
}
\end{figure}

We turn now to an intriguing scaling property exhibited by the
longitudinal resistance \Rhard. Figures 4a and 4b display the same \Rhard\ data
plotted against the {\it scaled} temperature $T/$\Bpar\ (for non-zero
\Bpar). Plotted in this way, the data 
collapse onto a single curve with impressive precision. Figs.~4c and 4d
show log-log plots of the computed resistivities \rhohard\ $vs.$
$T/$\Bpar\ and again the scaling is evident. This result demonstrates
that the hard resisitivities depend on temperature and in-plane field
only through the ratio $T/$\Bpar. The log-log plots reveal that this
dependence is quasi-linear: The dashed lines in the figure are
proportional to $(T/$\Bpar$)^{-0.75}$. We note in passing that the easy
direction resistance \Reasy\ does not exhibit this scaling behavior.
Although the origin of this difference is not known, it may be related
to the aforementioned current channeling effects which are inherent to
transport in anisotropic conductors.

The connection between the scaling behavior of \Rhard\ and the
nematic-to-isotropic phase transition picture is not yet understood.
While at sufficiently high temperature $T$ solutions to the 2D XY model
must eventually approach a Curie law $M\sim h/T$, 
this limiting behavior is not well developed in the ($h,T$)
parameter range which seems most relevant to our experimental data.
On the other hand, a quantitative comparison of the 2D XY model with
our data is hindered by several factors.
For example, the relation between the
symmetry-breaking parameter $h$ and the in-plane magnetic field \Bpar\
is subtle and highly sample-specific\cite{jungwirth}.  Also, \Bpar\
may affect the transport mechanism of current along and across the
stripes to some degree.
Even the appropriateness of the classical 2D XY model itself can be questioned.
We emphasize, however, that the experimental observation of scaling is robust and
offers a new insight into the nature of the anisotropic phases.

In the picture outlined above, the essential role of the in-plane
magnetic field is to orient previously existing local striped regions.
It is, however, also possible that \Bpar\ so alters the energetics
of the electron system that it {\it creates} stripe order at
temperatures far above where anisotropic transport appears spontaneously. 
If this is the
case, then the nematic-to-isotropic transition interpretation of our
data may be unjustified. Although we cannot completely rule out such a
scenario, there are good reasons to discount it. First, the data shown
above clearly indicate that the in-plane field severely broadens the
onset of resistive anisotropy; it does not merely shift the sharp
transition encountered at \Bpar\ $=0$ to a higher temperature. Second,
the scaling behavior we observe is sub-linear in $T/$\Bpar. A model in
which the both stripe formation and orientation are dependent on
\Bpar\ would presumably lead to a super-linear dependence on \Bpar.

In conclusion, we have found the anisotropy characteristic of transport
in half-filled high Landau levels to be heavily broadened by a
strong in-plane magnetic field. Although we have concentrated here on the
$\nu=9/2$ state, the same effect is observed at $\nu=11/2$ and other
half-filled high Landau levels. 
This finding suggests that local stripe
moments exist at relatively high temperatures where the transport is
isotropic in the absence of the in-plane field. In addition, we have
also observed a remarkable scaling of the resisitive anisotropy with the
ratio $T/$\Bpar. 

We thank E. Fradkin, S. Kivelson, and V. Oganesyan for helpful discussions.
This work was supported by the DOE under Grant No. DE-FG03-99ER45766 and
the NSF under Grant No. DMR0070890.


\begin{references}
\bibitem{bpar} Thermal cycling to room temperature was required to switch the 
direction of the parallel magnetic field between \ydir\ and \xdir.
\end{references}
\end{document}